\documentstyle[12pt,epsf]{article}
\newcommand{\rp}{\mbox{p}}
\newcommand{\rk}{\mbox{k}}
\newcommand{\rps}{\mbox{\footnotesize p}}
\newcommand{\ha}{\frac{1}{2}}
\newcommand{\refc}[1]{Ref.~\cite{#1}}

\begin{document}\baselineskip .7cm

\title{Relativistic Unitarized Quark/Meson Model \\ in Momentum Space} 
\author{
George Rupp$^{\,a}$ and Eef van Beveren$^{\,b}$ \\[5mm]
$^{a}${\footnotesize\it Centro de F\'{\i}sica das Interac\c{c}\~{o}es
Fundamentais, Instituto Superior T\'{e}cnico, P-1049-001 Lisboa, Portugal} \\
{\footnotesize\tt george@ajax.ist.utl.pt} \\[3mm]
$^{b}${\footnotesize\it Departamento de F\'{\i}sica, Universidade de Coimbra,
P-3004-516 Coimbra, Portugal} \\ {\footnotesize\tt eef@teor.fis.uc.pt} \\[5mm]
{\small hep-ph/0212100}
}
\date{\today}

\maketitle

\begin{abstract}
An outline is given how to formulate a relativistic unitarized constituent
quark model of mesons in momentum space, employing harmonic quark confinement.
As a first step, the momentum-space harmonic-oscillator potential is solved 
in a relativistically covariant, three-dimensional quasipotential framework for
scalar particles, using the spline technique. Then, an illustrative toy model
with the same dynamical equations but now one $q\bar{q}$ and one meson-meson
channel, coupled to one another through quark exchange describing the $^3P_0$
mechanism, is solved in closed form on a spline basis. Conclusions are
presented on how to generalize the latter to a realistic multichannel
quark/meson model.
\end{abstract}

\section{INTRODUCTION}
Mesonic $q\bar{q}$ states are the simplest color-singlet configurations in
QCD. However, physical mesons appear to be much more complicated objects, in
view of the quite disparate spectra of especially the light scalars and
pseudoscalars, and the large variety of states ranging from OZI-stable
quarkonia to extremely broad, often badly established resonances. Because of
these complexities, theoretical approaches usually focus on parts of the
meson data only. Thus, one has quarkonium potential models (see e.g.\
\refc{quarkonium}), relativistic constituent quark models (see e.g.\
\refc{rcqm}), chiral quark models (see e.g.\ \refc{cqm}), heavy-quark
effective theory \cite{hqet}, the linear $\sigma$ model (L$\sigma$M) \cite{lsm},
Nambu--Jona-Lasinio models \cite{njl}, (unitarized) chiral perturbation
theory \cite{chpt}), meson-exchange models (see e.g.\ \refc{mem}), and so on.
While it lies outside the scope of this short paper
to describe in detail the respective merits and drawbacks of all these
approaches, on can generally state that a good reproduction of either spectra
or scattering data is achieved, often at the cost of a non-negligible number of
free fit parameters, but without accomplishing a unified description of mesonic
spectra and meson-meson scattering. In our view, however, spectra and
scattering of mesons are inexorably intertwined concepts, considering that
most mesons are resonances, decaying strongly into final states of two or more
lighter mesons.

Therefore, we believe one should treat the mechanisms responsible for quark
confinement and for mesonic decay on an equal footing. This we have realized
already in an essentially non-relativistic (NR) framework, employing the
coupled-channel Schr\"{o}dinger formalism, with several confined $q\bar{q}$
channels and many free two-meson channels. For a detailed discussion of the
configuration-space model, we refer to a separate contribution \cite{eef02} 
to this workshop, so let us just emphasize here some important non-perturbative
features. From the spectroscopic point of view, meson-loop effects on the
$q\bar{q}$ spectra are included to all (ladder) orders, even those resulting
from closed meson-meson channels. Conversely, the strong influence of
$s$-channel $q\bar{q}$ states on non-exotic meson-meson scattering is taken
into account. Furthermore, bound states, resonances, and scattering observables
all follow directly from an analytic $S$-matrix derived in closed form, thus
dispensing with any kind of perturbative methods, which could be highly 
untrustworthy for the often very large couplings involved. As a matter of fact,
perhaps the most remarkable conquest of the model is its parameter-free
prediction of the light scalar mesons \cite{BRMDRR86}, including the
$\kappa$(800) ($K_0^*$) \cite{BR99}, which has very recently been experimentally
confirmed \cite{A02}. Precisely for the light scalars, perturbation theory
manifestly breaks down \cite{eef02,BR01}.

Other coupled-channel, also called unitarized, approaches are due to Eichten
{\em et al.} (\refc{quarkonium}, first paper), Bicudo \& Ribeiro (\refc{cqm},
fourth paper), and T\"{o}rnqvist with collaborators (\refc{TR96} and references
therein). The conclusions of the latter two models are in qualitative agreement
with ours, in the sense that large shifts are found owing to unitarization, even
for (bare) states below the lowest decay threshold. However, in the case of the
scalar mesons, important differences between our model and especially the one of
\refc{TR96} come to light, as the $\kappa$(800) is not found in the latter
reference (see \refc{BR99} for a detailed comparison).

Notwithstanding its successes, our coordinate-space model has, of course, a
number of shortcomings. First of all, the light pseudoscalar mesons are quite
badly off, except for the kaon. This may be due to the largely NR formalism
without Dirac spin structure , the disregard of chiral symmetry, and possibly
also the neglect of scalar-pseudoscalar two-meson channels. Moreover, for most
processes, the experimental meson-meson phase shifts are only roughly and not
very accurately reproduced, which is no wonder in view of the neglect of
final-state interactions from direct $t$-channel meson exchange, and the
absence of any ``fitting freedom''. Then there are problems with
pseudothresholds in heavy-light systems, owing to the non-covariance of 
minimally relativistic equations in configuration space.
Finally, too many, viz.\ 3,  parameters  --- albeit a very modest number as
compared to other approaches --- are needed for the $^3P_0$ transition
potential, due to the lack of a microscopic description in terms of quark
exchange.

Nevertheless, we are convinced one can deal with these shortcomings
by reformulating the model in momentum space, in a covariant three-dimensional
(3D) quasipotential framework. Here, one could object why we do not attempt
to tackle the full Bethe-Salpeter (BS) equation. The reason is twofold: first
of all, the numerical effort to solve the resulting coupled 4D
integro-differential equations with many channels would be enormous; but
perhaps more importantly, the confinement mechanism employing potentials is an
inherently 3D concept, possibly leading to a pathological, non-confining
behavior when generalized to four dimensions (see e.g.\ \refc{rcqm}, third
paper). Anyhow, a covariant formulation would also allow to include, at a
later stage, the Dirac spin structure of the quarks, and address the issue of
dynamical chiral-symmetry breaking and generation of constituent mass. The
resulting multichannel equations in momentum space we shall show to be
explicitly solvable, provided we stick to harmonic confinement and work on a
spline basis.

This paper is organized as follows. In Sec.~2 the good old harmonic oscillator
(HO) is solved in momentum space using the spline technique, for two different
3D relativistic equations. In Sec.~3 a two-channel toy model, with one
harmonically confined $q\bar{q}$ channel coupled via quark exchange to one free
two-meson channel, is formally solved in closed form, employing again a spline
expansion. Conclusions and an outlook on how to further develop the model are
presented in Sec.~4.

\section{RELATIVISTIC HARMONIC OSCILLATOR}
In this section, we review the well-known 3D HO, first in coordinate space,
and then in momentum space. In the latter representation, a 3D relativistic
generalization is straightforward.
\subsection{Non-Relativistic Harmonic Oscillator}
The equal-mass radial Schr\"{o}dinger equation (SE) for a local potential reads
\begin{equation}
\left\{\frac{1}{m}(-\frac{d^2}{dr^2}+\frac{l(l+1)}{r^2})+V(r)\right\}u(r)
\; = \; Eu(r) \;.
\end{equation}
If we take the HO potential
\begin{equation}
V(r) \; = \; \frac{1}{2}\mu\,\omega^2r^2\;,\;\;\;\mu \; = \; \frac{1}{2}m\,,
\end{equation}
define $x\equiv\sqrt{\mu\omega}\,r$, and assume for the moment $l=0$, we get
\begin{equation}
\left\{-\frac{d^2}{dx^2}+x^2\right\}u(x) \; = \; \frac{E}{\frac{1}{2}\omega}\
,u(x)\;.
\end{equation}
The spectrum of the 3D HO potential is very well known, viz.\
($l=0$)
\begin{equation}
E \; = \; \omega(2n+\frac{3}{2})\;,\;\;\;n=0,1,2,\ldots\,,
\end{equation}
leading to the eigenvalues $3,7,11,\ldots$ for Eq.~(3).

Alternatively, we can work in momentum space. The homogeneous
Lippmann-Schwinger (LS) equation  for the bound-state vertex function reads
\begin{equation}
\Gamma(\vec{p}) \; = \; \int \frac{d^3p'}{(2\pi)^3}\,V(\vec{p},\vec{p}\,')\,
G_0(\vec{p}\,';E)\,\Gamma(\vec{p}\,') \;,
\end{equation}
where the free two-body Green's function is given by
\begin{equation}
G_0^{-1}(\vec{p}\,') \; = \; E-\frac{\vec{p}\,'^2}{m}+i\epsilon\;.
\end{equation}
Now define the wave function $\Psi\equiv G_0\,\Gamma$. Then Eq.~(5) becomes
\begin{equation}
G_0^{-1}(\vec{p})\Psi(\vec{p}) \; = \; \int \frac{d^3p'}{(2\pi)^3}\,
V(\vec{p},\vec{p}\,') \, \Psi(\vec{p}\,') \;,
\end{equation}
or equivalently
\begin{equation}
\frac{\vec{p}\,^2}{m}\,\Psi(\vec{p})+\int \frac{d^3p'}{(2\pi)^3}\,V(\vec{p},
\vec{p}\,')\,\Psi(\vec{p}\,') \; = \; E\,\Psi(\vec{p}) \;.
\end{equation}
This is just the SE in momentum space. In this representation, the
HO potential is given by
\begin{equation}
V(\vec{p},\vec{p}\,') \; = \; -\frac{1}{2}\mu\omega^2\,\delta^{(3)}
(\vec{p}-\vec{p}\,')\,\Delta_{\vec{p}} \;,
\end{equation}
which is local as in coordinate space.
Substituting into (8) yields
\begin{equation}
\left\{\frac{\vec{p}\,^2}{m}\,-\,\frac{1}{2}\mu\omega^2\,
\Delta_{\vec{p}}\right\}\Psi(\vec{p}) \; = \; E\,\Psi(\vec{p}) \;.
\end{equation}
Taking again $l=0$ and introducing $v(\rp)=\rp\,\Psi(\rp)$, with
$\rp\equiv|\vec{p}|$, we obtain
\begin{equation}
\left\{\frac{\rp^2}{m}\,-\,\frac{1}{2}\mu\omega^2\frac{d^2}{d\rp^2}\right\}
v(\rp) \; = \; E\,v(\rp) \;.
\end{equation}
If we define the dimensionless variable $y\equiv\rp/\sqrt{\mu\omega}$ and
write $m=2\mu$, we get
\begin{equation}
\left\{y^2\,-\,\frac{d^2}{dy^2}\right\}v(y) \; = \;
\frac{E}{\frac{1}{2}\omega}\,v(y) \;.
\end{equation}
This is identical to Eq.~(3). 
Note that now the roles of the kinetic and the potential term are interchanged.
\subsection{Blankenbecler--Sugar--Logunov--Tavkhelidze (BSLT) Equation}
The momentum-space equal-mass (scalar) BSLT \cite{bslt} equation for the wave
function has exactly the same form as in the LS case of Eq.~(7), but now with
center-of-mass (CM) Green's function 
\begin{equation}
G_0^{-1}(\vec{p}) \; = \; \frac{E_p}{m^2}\,(\,\frac{s}{4}-E_p^2+i\epsilon)\;,
\end{equation}
where $E_p=\sqrt{\vec{p}\,^2+m^2}$, and $s$ is the total CM energy squared
$s=P^2=(p_1+p_2)^2=(2m+E)^2$. 
Note that $G_0^{\mbox{\footnotesize BSLT}}$ can be derived in a completely
covariant fashion, and written down explicitly for an arbitrary frame
\cite{WJ73}.
In the NR limit we have $E_p\approx m$ and $s\approx4m^2+4mE$, recovering the
LS Green's function (6). With the new $G_0^{-1}$, we get instead of
Eq.~(8)
\begin{equation}
\frac{\vec{p}\,^2}{m}\,\Psi(\vec{p})\,+\,\frac{m}{E_p}\int
\frac{d^3p'}{(2\pi)^3}\,V(\vec{p},\vec{p}\,')\,\Psi(\vec{p}\,') \; = \; 
\frac{s-4m^2}{4m}\:\Psi(\vec{p}) \;.
\end{equation}
In terms of the above-defined dimensionless variable $y$, the BSLT equation
becomes
\begin{equation}
\left\{y^2\,-\,\frac{1}{\displaystyle\sqrt{1+\frac{\omega}{2m}y^2}}\,
\frac{d^2}{dy^2}\right\}v(y) \; = \; \frac{s-4m^2}{2m\omega}\:v(y) \,.
\end{equation}
From this equation we readily see that deviations from the LS case are 
determined by the parameter $\omega/2m$. Furthermore, in the limit
$\omega/2m\rightarrow0$, the eigenvalues $3,7,11,\ldots$ get recovered.
However, there are additional relativistic effects due to $s\neq4m^2+4mE$.
\subsection{Equal-Time (ET) Equation}
The ET, also called Mandelzweig--Wallace \cite{WM89}, Green's function can be
obtained from the BSLT (or Salpeter) equation by
adding an extra term $G_0^C$ to the two-body Green's function $G_0$ which
approximates crossed-ladder contributions. This extra term becomes exact
in the eikonal limit, or when one of the particles gets infinitely heavy.
Therefore, the covariant ET equation manifestly has the correct one-body limit,
i.e., one recovers either the Klein--Gordon or the Dirac equation, for scalar
or spin-\ref{relho}/2 particles, respectively.
In the scalar case, the CM ET Green's function is given by
\begin{equation}
G_0^{-1}(\vec{p}) \; =\; \frac{E_p^3}{m^2}\;\frac{\displaystyle\frac{s}{4}-
E_p^2+i\epsilon}{2E_p^2-\displaystyle\frac{s}{4}}\;.
\end{equation}
The final form of the $S$-wave ET equation for the HO potential in momentum
space reads
\begin{equation} 
\left\{y^2\,-\,\frac{1}{\displaystyle\sqrt{1+\frac{\omega}{2m}y^2}}
\left[2-\frac{1}{\displaystyle\displaystyle1+\frac{\omega}{2m}y^2}\right]
\frac{d^2}{dy^2}\right\} v(y) \; = \; \frac{s-4m^2}{2m\omega}\:
\left\{1\,-\,\frac{\displaystyle\frac{\omega}{2m}}{\left(\displaystyle1+
\frac{\omega}{2m}\right)^{\frac{3}{2}}}\, \frac{d^2}{dy^2} \right\}v(y) \;.
\end{equation}
\subsection{Non-Zero Orbital Angular Momentum}
For $l\neq0$, the dimensionless LS equation becomes
\begin{equation}
\left\{-\frac{d^2}{dx^2}\,+\,\frac{l(l+1)}{x^2}\,+\,x^2\right\}u(x) \; = \;
\frac{E}{\frac{1}{2}\omega}\,u(x)\;.
\end{equation}
In order to numerically solve the LS and corresponding BSLT, ET equations,
we substitute
$u(x)\rightarrow x^l\tilde{u}(x)$. Then
\begin{equation}
\left\{-\frac{d^2}{dx^2}\,+\,\frac{2l}{x^2}(1-x\frac{d}{dx})\,+\,x^2\right\}
\tilde{u}(x) \; = \; \frac{E}{\frac{1}{2}\omega}\,\tilde{u}(x)\;.
\end{equation}
The function $\tilde{u}(x)$ has the right boundary conditions to allow an easy
numerical solution using an expansion on a spline basis (for details, see
\refc{P87} and second paper of \refc{rcqm}).
Thus, the NR HO spectrum is recovered with great accuracy, viz.\
\begin{equation}
E \; = \; \omega\,(2n\,+\,l\,+\,\frac{3}{2}) \;, \, \;\;\;n,l=0,1,2,\ldots \; . 
\end{equation}
\mbox{ } \\[-1cm]
\begin{figure}
\centerline{\epsffile{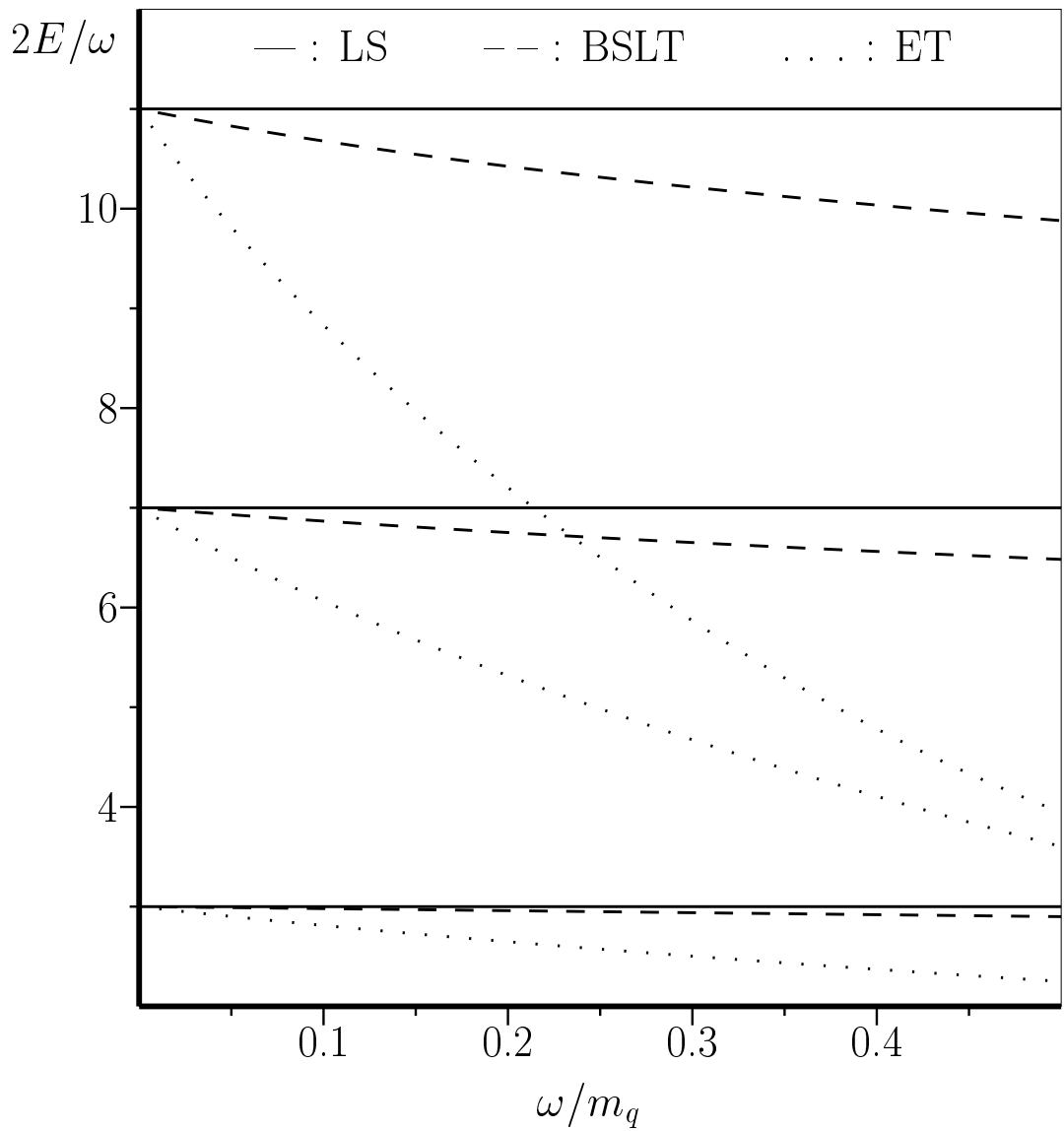}}
\caption{Dimensionless energy levels of the HO for the LS, BSLT, and ET
equations, as a function of $\omega/m_q$; first three radial levels are
depicted ($l=0$).}
\label{relho}
\end{figure}

\noindent In the case of the BSLT and ET  equations, the same can be done by
replacing everywhere
\begin{equation}
-\frac{d^2}{dy^2} \; \longrightarrow \; \frac{2l}{y^2}\,(1-y\frac{d}{dy})
\, - \, \frac{d^2}{dy^2} \; .
\end{equation}
The LS, BSLT, and ET results for $l=0$ are depicted in Fig.~\ref{relho}. We see
that the relativistic corrections are relatively modest for the BSLT
equation, but can become huge in the ET case, especially for excited states.
The latter large effect has to do with the sign change of the ET propagator
(16) as a function of the relative momentum, and has also been observed in the
Dirac case (see \refc{rcqm}, second paper). However, this overshooting of the
ET equation will strongly depend on the chosen Dirac structure of the confining
force in a realistic model, and hence should not cause any worries at this
stage.
\section{UNITARIZED MODEL IN MOMENTUM SPACE}
\begin{figure}
\centerline{\epsffile{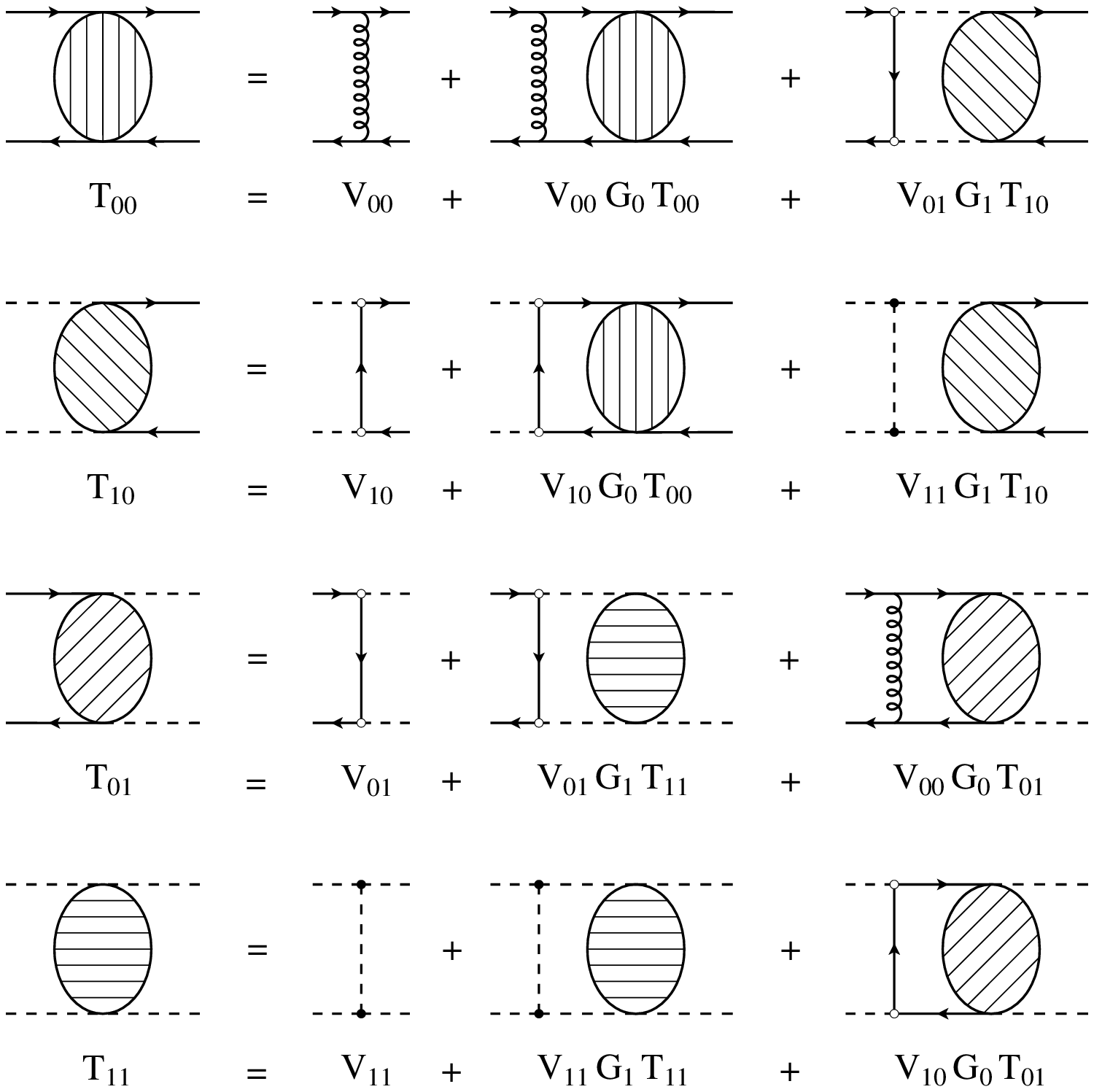}}
\caption{Graphical representation of Eq.~(22); see text for explanation.}
\label{tij}
\end{figure}

Consider now a simple two-channel model of one confined $q\bar{q}$ system
coupled to one free meson-meson channel. In a unitarized picture, this can
formally be described by a $2\!\times\!2\,$ $T$ matrix, with the dynamical
equation (LS, BSLT, or ET)
\begin{equation}
\left( \begin{array}{cc}
T_{00} & T_{01} \\ T_{10} & T_{11}
\end{array} \right) \; = \; 
\left( \begin{array}{cc}
V_{00} & V_{01} \\ V_{10} & V_{11}
\end{array} \right) \: + \: 
\left( \begin{array}{cc}
V_{00} & V_{01} \\ V_{10} & V_{11}
\end{array} \right) 
\left( \begin{array}{cc}
G_{0} & 0 \\ 0 & G_{1}
\end{array} \right) 
\left( \begin{array}{cc}
T_{00} & T_{01} \\ T_{10} & T_{11}
\end{array} \right) \; .
\end{equation}
Equation (22) is diagrammatically represented in Fig.~\ref{tij}.
Dashed lines are mesons, solid lines with arrows are (anti)quarks, and
curly exchanges symbolize confinement. Furthermore,
open circles  represent meson-quark-antiquark vertices, and dots stand for
three-meson vertices. Note that the communication between the $q\bar{q}$ and
two-meson channels takes place through quark exchange. This will give rise
to a generally non-local $^3P_0$ transition potential in configuration space,
as soon as the quarks and mesons have different masses. The precise form of
such a potential will also depend on possible form factors to be attributed
to the various vertices.

In order to solve Eq.~(22), we must realize that no asymptotic scattering of
quarks is possible, so $T_{00}$, $T_{01}$, and $T_{10}$ must in the end be
eliminated from the equations: only $T_{11}$ can be on shell.
If we now take again harmonic confinement, Eq.~(22) becomes a set of coupled
integro-differential equations. Omitting for clarity angular-momentum indices,
we get e.g.\ for $T_{01}$ 
\begin{equation}
T_{01}(\rp,\rp') \; = \; V_{01}(\rp,\rp')\,-\,\ha\mu\omega^2
\Delta_{\rps}\,G_0(\rp)\,T_{01}(\rp,\rp') \,+\,
\int\frac{\rk^2d\rk}{4\pi^2}\,V_{01}(\rp,\rk)\,G_1(\rk)\,T_{11}(\rk,\rp') \; .
\end{equation}
Note that $\rp'$ is an essentially dummy variable in the equations, so
we can choose a fixed value for it, say $\bar{\rp}$. If we now expand
$T_{01}$ and $T_{11}$ on a spline basis and discretize the initial momentum
$\rp$, then $T_{01}$ can be expressed in terms of $T_{11}$ as follows.
\begin{equation}
T_{01}(\rp_i,\bar{\rp}) \; = \; \sum_{j=1}^{N}t_{01}^{j}s_{01}^{j}(\rp_i)
\; , \;\;\; v_{01}^i\;\equiv\;V_{01}(\rp_i,\bar{\rp}) \; ;
\end{equation}
\mbox{ } \\[-9mm]
\begin{equation}
T_{11}(\rk,\bar{\rp}) \; = \; \sum_{j=1}^{N}t_{11}^{j}s_{11}^{j}(\rk) \; ;
\end{equation}
\begin{equation}
A_{01}^{ij} \; \equiv \; \left\{1\,+\,\ha\mu\omega^2\,
\Delta_{\rps_i}\,G_0(\rp_i)\right\} s_{01}^j(\rp_i) \; ; 
\end{equation}
\begin{equation}
A_{11}^{ij} \; \equiv \; \int\frac{\rk^2d\rk}{4\pi^2}\,V_{01}(\rp_i,\rk)\,
G_1(\rk)\,s_{11}^j(\rk)\; .
\end{equation}
Combining Eqs.~(24)--(27), we can write
\begin{equation}
{\bf A}_{01}\cdot{\bf t}_{01}\;=\;{\bf v}_{01}\,+\,{\bf A}_{11}\cdot
{\bf t}_{11}\;\;\Rightarrow\;\;{\bf t}_{01}\;=\;{\bf A}_{01}^{-1}\cdot
\{{\bf v}_{01}\,+\,{\bf A}_{11}\cdot{\bf t}_{11}\} \; ,
\end{equation}
where the boldface capitals and small letters  are obvious shorthands for
matrices and vectors, respectively. Turning next to
$T_{11}$, we get the integral equation
\begin{equation}
T_{11}(\rp,\bar{\rp}) \; = \; V_{11}(\rp,\bar{\rp})\,+\,\int\frac{\rk^2d\rk}
{4\pi^2}\,V_{11}(\rp,\rk)\,G_1(\rk)\,T_{11}(\rk,\bar{\rp}) \, + \, \int
\frac{\rk^2d\rk}{4\pi^2}\,V_{10}(\rp,\rk)\,G_0(\rk)\,T_{01}(\rk,\bar{\rp})\;.
\end{equation}
Similarly to Eqs.~(24)--(27), we define now
\begin{equation}
B_{11}^{ij} \; \equiv \; s_{11}^j(\rp_i) \, - \, \int\frac{\rk^2d\rk}{4\pi^2}
\,V_{11}(\rp_i,\rk)\,G_1(\rk)s_{11}^j(\rk)\;;
\end{equation}
\begin{equation}
B_{01}^{ij}\;\equiv\;\int\frac{\rk^2d\rk}{4\pi^2}\,V_{10}(\rp_i,\rk)\,G_0(\rk)
\,s_{01}^j(\rk)\; , \;\;\; v_{11}^i \;\equiv\;V_{11}(\rp_i,\bar{\rp}) \; .
\end{equation}
Then we can express ${\bf t}_{11}$ in terms of itself, using Eq.~(28): \\[-2mm]
\begin{equation}
{\bf B}_{11}\cdot{\bf t}_{11}\;=\;{\bf v}_{11}\,+\,{\bf B}_{01}\cdot
{\bf t}_{01}\;\;\raisebox{-3mm}{$\stackrel{\displaystyle=}{\scriptstyle(28)}$}
\;\;{\bf v}_{11}\,+\,{\bf B}_{01}\cdot{\bf A}_{01}^{-1}\cdot\{{\bf v}_{01}\,+\,
{\bf A}_{11}\cdot{\bf t}_{11}\} \; .
\end{equation}
\mbox{ } \\[-5mm]
This finally leads to
\begin{equation}
\{{\bf B}_{11}\,-\,{\bf B}_{01}\cdot{\bf A}_{01}^{-1}\cdot{\bf A}_{11}\}\cdot
{\bf t}_{11} \; = \; {\bf v}_{11}\,+\,{\bf B}_{01}\cdot{\bf A}_{01}^{-1}\cdot
{\bf v}_{01} \; ,
\end{equation}
\begin{equation}
{\bf t}_{11} \; = \; \{{\bf B}_{11}\,-\,{\bf B}_{01}\cdot{\bf A}_{01}^{-1}\cdot
{\bf A}_{11}\}^{-1}\:\cdot\:\{{\bf v}_{11}\,+\,{\bf B}_{01}\cdot
{\bf A}_{01}^{-1} \cdot{\bf v}_{01}\} \; .
\end{equation}

\mbox{ }

Thus, we have obtained a closed-form expression for the $T$ matrix (hence the
$S$ matrix), containing the {\em complete} \/information of the
confinement-plus-scattering process. Needless to say that all these algebraic
manipulations only make sense if the involved integrals actually exist and the
spline expansions converge. Yet, we are confident this will be the case when
appropriate boundary conditions are chosen, since the same spline expansion we
are using here has been successfully applied to few-body scattering
calculations in momentum space \cite{PGSO92}. Nevertheless, in the Dirac case,
to be dealt with in the future, form factors will have to be included for the
meson-quark-antiquark vertices so as to make the integrals involving three
quark propagators convergent. In any case, such form factors are physically
meaningful, and should therefore already be considered in the approximation
with spinless fermions, whenever doing phenomenology.

Bound states and resonances are given by zeroes in the determinant of the
inverted-matrix factor right after the equal sign in Eq.~(34). Resonance poles
in the second Riemann sheet can be searched for by analytic continuation to
complex energies,
using e.g.\ contour-rotation methods.  Moreover, from the structure of the
equations one readily sees that the inclusion of final-state interactions, due
to $t$-channel meson exchange in the meson-meson channel, simply amounts to
taking a non-vanishing $V_{11}$ in Eq.~(29), which will not lead to any
additional numerical effort. Finally, the generalization to many scattering
channels and also more $q\bar{q}$ channels is straightforward, at the expense
of bigger matrices.
\section{CONCLUSIONS AND OUTLOOK}
In the foregoing, we have demonstrated that the formulation of a unitarized
quark/meson model with harmonic confinement is relatively easy in momentum
space. Thus, the $^3P_0$ mechanism is dynamically described through $P$-wave
quark exchange.  Furthermore, the inclusion of final-state interactions via
$t$-channel meson exchange is straightforward, and does not complicate the
structure of the equations. The numerical resolution of these equations using
splines looks very convenient, which should result in (approximate) analytic
solutions, thereby facilitating an analytic continuation to complex energies 
in order to search for resonance poles.

Our first priority is, of course, to achieve a total control of the numerics,
for real as well as complex energy. Then a simple toy model will be worked out
so as to reproduce known coordinate-space results. Furthermore, cases
problematic in the configuation-space formulation will be studied, such as
heavy-light mesons. Also, the influence of
$t$-channel meson exchange on meson-meson phase shifts will be investigated.
In a somewhat more distant perspective, we intend to do a lot of phenomenology,
and use the feedback to further refine the model. \\[3mm]

\noindent{\large\bf Acknowledgments}
The authors thank F.~Kleefeld and J.~A.~Tjon for valuable discussions, and
P.~Nogueira for expert graphical assistance.
This work was partly supported by the
{\em Funda\c{c}\~{a}o para a Ci\^{e}ncia e a Tecnologia} (FCT) 
of the {\em Minist\'{e}rio do Ensino Superior, Ci\^{e}ncia e Tecnologia} of 
Portugal, under contract number CERN/\-P/\-FIS/\-43697/\-2001.

\end{document}